\documentclass[iicol,sn-mathphys-ay]{sn-jnl}

\usepackage{graphicx}%
\usepackage{multirow}%
\usepackage{amsmath,amssymb,amsfonts}%
\usepackage{amsthm}%
\usepackage{mathrsfs}%
\usepackage[title]{appendix}%
\usepackage{xcolor}%
\usepackage{textcomp}%
\usepackage{manyfoot}%
\usepackage{booktabs}%
\usepackage{algorithm}%
\usepackage{algorithmicx}%
\usepackage{algpseudocode}%
\usepackage{listings}%

\usepackage{bm}
\usepackage{graphicx}
\usepackage{subcaption}
\usepackage{acronym}
\usepackage{acronym}

\acrodef{AIE}[AIE]{adaptive input estimation}
\acrodef{AUV}[AUV]{autonomous underwater vehicle}
\acrodef{IMU}[IMU]{inertial measurement unit}
\acrodef{RCIE}[RCIE]{retrospective cost input estimation}
\acrodef{RLS}[RLS]{recursive least squares}
\acrodef{ASV}[ASV]{autonomous surface vehicle}

\graphicspath{{./Plots/}}

\theoremstyle{thmstyleone}%
\theoremstyle{thmstyletwo}%
\newtheorem{remark}{Remark}%

\theoremstyle{thmstylethree}%

\begin{document}

\title[Article Title]{System Identification and Adaptive Input Estimation on the Jaiabot Micro Autonomous Underwater Vehicle}


\author[1]{\fnm{Ioannis} \sur{Faros}}\email{ifaros@udel.edu}
\equalcont{These authors contributed equally to this work.}
\author[1]{\fnm{Herbert G.} \sur{Tanner}}\email{btanner@udel.edu}
\equalcont{These authors contributed equally to this work.}



\affil[1]{\orgdiv{Mechanical Engineering Department}, \orgname{University of Delaware}, \orgaddress{\street{130 Academy St}, \city{Newark}, \postcode{19716}, \state{DE}, \country{USA}}}



\abstract{This paper reports an attempt to model the system dynamics and estimate both the unknown internal control input and the state of a recently developed marine autonomous vehicle, the Jaiabot. Although the Jaiabot has shown promise in many applications, process and sensor noise necessitates state estimation and noise filtering. In this work, we present the first surge and heading linear dynamical model for Jaiabots derived from real data collected during field testing. An adaptive input estimation algorithm is implemented to accurately estimate the control input and hence the state. For validation, this approach is compared to the classical Kalman filter, highlighting its advantages in handling unknown control inputs.}




\maketitle

\section{Introduction}

\label{sec:introduction} 
Over the last decades, the effects of climate change, including rising sea levels, disruptions to marine life, and declining  water quality, have become increasingly prominent \citep{liu2016unmanned,zereik2018challenges}. As a result, the demand for accurate environmental monitoring and in situ measurements around coastal regions has grown significantly. 
To observe such phenomena and collect data through traditional methods, which primarily rely on in situ measurements from stationary platforms or manual data collection, can either yield very sparse data sets or be time consuming and costly. 
From this perspective, such data collection tasks can be serviced by automating these processes by using one or multiple \acp{AUV} \citep{lim2023applied,9123685,li2021auv}. These \acp{AUV} can be deployed from shore and offer advantages in terms of affordability and open source access.

\begin{figure}[h!]
    \centering  \includegraphics[width=0.95\columnwidth]{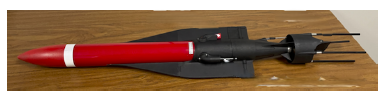}
    \caption{The first generation of the Jaiabot AUV.}
    \label{fig:Jaiabot}
\end{figure}

One such micro-\ac{AUV} that shows great promise for environmental monitoring application is the Jaiabot (Fig.~\ref{fig:Jaiabot}). 
This vehicle is approximately one meter long, weighs about 3 kg, and is capable of achieving speeds of nearly $5$ m/s with a range of 11 km. 
In order to perform vertical dives or general movement within the sea, the Jaiabot is equipped with a single propeller, rudder, GPS, and compass, \ac{IMU}, plus with sensors for collecting data such as salinity and temperature. 
Few attempts at constructing dynamical models for such micro-\acp{AUV} have been reported. 
The first known dynamical system for a Jaiabot is known for its vertical dive motion~\citep{tanner2024partial}, based on which, a control for diving is redesigned and developed with new safety constraints, specifically with respect to overshoot in terms of assigned depth. 
Subsequent design modifications made on this \ac{AUV} require new system parameter identification, while the aforementioned approach applies still.

While the controlling the Jaiabot during vertical dive without overshoot has been addressed, open questions still remain regarding its surface maneuvering. 
Thus in this paper, the Jaiabot is operated mainly as an \ac{ASV}.
Specifically, noisy data collected from the miniature onboard GPS and \ac{IMU} may result in inaccurate positioning. 
In classical state estimation, the Kalman filter and its variants give well-established techniques for estimating unmeasured states by taking advantage of knowledge about the system dynamics, its input, as well the process and sensor noise.
However, while the PID control loop that regulates thruster speed and rudder configuration of this \ac{ASV} is known, the  actual thrust generated to propel the \ac{ASV} or the torque generated to make it turn is unknown. 
This presents challenges to the implementation of a Kalman filter on the surface maneuvering dynamics. 
Specifically, here both the state of the system and its input have to be estimated simultaneously. 

This is not a new problem. 
Numerous methods have been used for state estimation in noisy environments \citep{paull2013auv}. 
However, when the input signal is deterministic but unknown, obtaining unbiased state estimates becomes crucial. 
Solutions to this problem include unbiased Kalman filters, unknown input observers, and sliding-mode observers \citep{kitanidis1987unbiased,darouach1997unbiased,veluvolu2009high}. 
An alternative input estimation approach considers the unknown input as the output of an auxiliary system with known dynamics, perturbed by white noise. 
This estimated input can then be incorporated into the state estimator to yield more accurate state estimates. Over the last decade, a new technique known as \ac{RCIE} has been developed \citep{ansari2018input,sanjeevini2022decomposition}. 
\ac{RCIE} formulates a \emph{retrospective cost optimization} problem, where the coefficients of the input estimator are recursively adjusted to minimize a (retrospective) cost function. 
By doing so, \ac{RCIE} effectively builds an internal model of the unknown input that estimates the later, which is subsequently fed into the Kalman filter. 
The estimator coefficients are continuously adapted using the innovations (differences between predicted and observed measurements) as the error metric, thereby enhancing the accuracy and robustness of state estimation. \ac{RCIE} has been extensively studied and modified for nonminimum-phase discrete time systems, linear time varying \citep{SANJEEVINI2024105744} and invariant systems \citep{7525440}.
It has also been applied in the area of signal and process and especially in numerical differentiation \citep{verma2024real} and integration \citep{10155990}. 


The \textit{contributions} of the work reported in this paper are outlined as follows: 
\begin{itemize}
    \item \textit{Modeling surge and heading dynamics}: The surge and heading dynamics of Jaibot are derived from field data.
    \item \textit{\Ac{AIE}}: The adaptive input estimation based on the \ac{RCIE} \citep{verma2024real}, is implemented to the derived models of Jaiabot to estimate the internal control input and hence the state of the system using real data.
\end{itemize}
The remaining of the paper is organized as follows. Section \ref{section_2} frames the process of modeling the surge and heading dynamics from real data. Section \ref{validation}
provides the simulation results using real data in support of the derived models, the implemented AIE algorithm, and give a comparative analysis between AIE and the classical Kalman filter. Section \ref{conc} summarizes the research outcomes and outlines directions for future work.

\section{Technical Approach}\label{section_2}

\subsection{System Identification} 

The Jaiabot is a miniature marine vehicle that can move fast on the surface of the water.
To perform principled control design for automated maneuvering for this vehicle we constructed models based on experimental data.
There can be several general model templates for surface vehicle kinematics and dynamics \citep{Fossen}, and most advanced such models incorporate coupling between longitudinal (surge) and lateral (yaw) vehicle motions \citep{JDSMC15}.
Given the choice of parameterizing the model based on \emph{experimental data}, we opted to initially ignore the coupling between surge and yaw (see also \citep{Fossen}) and built separate models for surge and yaw motion.

\subsubsection{Surge dynamics}
\begin{figure}[h!]
\centering
\begin{subfigure}[b]{0.2\textwidth}
\centering
\includegraphics[width=\textwidth]{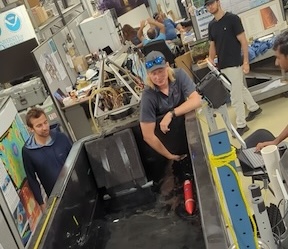}
\caption{Surge tests}\label{figure:bootcamp-surge}
\end{subfigure}
\begin{subfigure}[b]{0.26\textwidth}
\centering
\includegraphics[width=\textwidth]{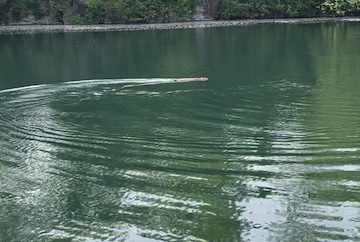}
\caption{Yaw tests}\label{figure:Allure-lateral}
\end{subfigure}
\caption{Maneuvering experiments for motion dynamics system identification. 
(\subref{figure:bootcamp-surge})
Experimental data for surge dynamics were obtained from indoor tests in a long water tank.
(\subref{figure:Allure-lateral})
Data for yaw dynamics were collected during outdoor tests in lake Allure, PA
}\label{figure:SYSID}
\end{figure}

The surge dynamics of the vehicle were assumed \citep{Fossen} to take the form of a second order linear system
\begin{equation} \label{surge model}
m \,\ddot{x} + d \,\dot{x} = u
\end{equation}
where $m$ denotes the mass of the vehicle, $d$ is in the role of a hydrodynamic drag coefficient along the surge direction, and $u$ expresses forward thrust.
Equivalently, \eqref{surge model} can be regarded as a first order system in surge speed; either way, the step response of this model is known analytically and thus knowing the mass of the vehicle, the coefficient $d$ can be directly determined through a least squares process from experimental data (Fig.~\ref{figure:bootcamp-surge})
The fitting process, accounting for input scaling, yields parameter values for \eqref{surge model} as follows: $m = 0.469$, and $d = 0.311$.

\subsubsection{Heading dynamics}
The identification of the heading (yaw) dynamics of the Jaiabot was performed using a standard maneuvering test for marine vehicles \citep{Fossen}.
The vehicle moves with a fixed rudder configuration over a period of time (Fig.~\ref{figure:Allure-lateral}), and the response of the yaw rate is measured (in this case, based on the vehicle's \ac{IMU} and GPS readings).
Assuming a similar second order model for the yaw dynamics 
\begin{equation} \label{yaw model}
I \, \ddot{\theta} + c \,\dot{\theta} = r
\end{equation}
where $I$ is in the role of the vehicle's moment of inertia along the vertical axis, $c$ is a hydrodynamic drag coefficient and $r$ expresses the rudder input command, a least squares fitting approach on the step response for this model gives $I = 4.896 $ and $c = 9.087$.

In the section that follows, the surge and yaw dynamics, \eqref{surge model} and \eqref{yaw model} respectively, are expressed in discrete time, in order to be integrated into a Kalman filter.

\subsection{Discrete-time Vehicle Dynamics}

Jaiabot's motion, just like most surface vehicles, is along curves; neither straight lines nor pure rotations.
Thus none of the models of \eqref{surge model} or \eqref{yaw model} can, in isolation, describe the motion of the vehicle.
What is more, to be integrated into a Kalman filter, these models need to be associated with measurable outputs.
And while the vehicle's orientation is directly measured via its \ac{IMU} or compass, the a single GPS measurement does not directly inform about the length of the path traveled, which is practically what is being tracked by \eqref{surge model}.
To overcome the latter challenge we construct discrete-time models that describe the evolution of path length traversed by the vehicle and its change in orientation between two consecutive \emph{time steps}.

\subsubsection{Heading}
We start with the incremental heading dynamics because they are more straightforward.
If we name $z_1$ a variable capturing the vehicle's current heading and $z_2$ its yaw rate, then the discrete-time heading dynamics derived by analytic integration of \eqref{yaw model} for a time step $T$\footnote{In implementation, $T$ is set to $0.546$ seconds.} results in 
\begin{multline*}
\left[\begin{smallmatrix}
z_1 \\ z_2 
\end{smallmatrix}\right]_{k+1}
=
\left[\begin{smallmatrix}
1 & 0.5487(1 - \mathrm{e}^{-1.82249T})  \\ 
0 & \mathrm{e}^{-1.82249T} 
\end{smallmatrix}\right]
\left[\begin{smallmatrix}
z_1 \\ z_2 
\end{smallmatrix}\right]_{k}
\\ +
\left[\begin{smallmatrix}
0.5487 + 0.301072(\mathrm{e}^{-1.82249T}) \\ 
0.5487(1 - \mathrm{e}^{-1.82249T}) \\ 
\end{smallmatrix}\right] r_k
\end{multline*}

We now augment the heading state vector at step $k+1$ to include the heading at step $k$ as so
\begin{multline}
\left[\begin{smallmatrix}
z_1 \\ z_2 \\ z_3 
\end{smallmatrix}\right]_{k+1}
=
\left[\begin{smallmatrix}
1 & 0.5487(1 - \mathrm{e}^{-1.82249T}) & 0  \\ 
0 & \mathrm{e}^{-1.82249T} & 0 \\ 
1 & 0 & 0  
\end{smallmatrix}\right]
\left[\begin{smallmatrix}
z_1 \\ z_2 \\ z_3 
\end{smallmatrix}\right]_{k}
\\ +
\left[\begin{smallmatrix}
0.5487 + 0.301072(\mathrm{e}^{-1.82249T}) \\ 
0.5487(1 - \mathrm{e}^{-1.82249T}) \\ 0 
\end{smallmatrix}\right] r_k
\label{heading-discrete-time}
\end{multline}
and define the output of this discrete-time linear system to be $\Delta\bar{\theta}$ which relates to the heading state $z$ through
\begin{equation}
\label{incremental-heading}
\Delta\bar{\theta}_{k+1} 
= 
0.200474 \left[\begin{smallmatrix}
1 & 0 & -1 
\end{smallmatrix}\right]
\left[\begin{smallmatrix}
z_1 \\ z_2 \\ z_3 
\end{smallmatrix} \right]_{k+1}
\end{equation}
The output variable can be considered directly measurable in the form of the difference of compass readings between two time steps.

\subsubsection{Surge} 
The discrete-time augmented dynamics for surge is constructed along the same lines. 
We denote $z_1$ the length of the path traveled by the vehicle, $z_2$ its speed along this path, and $z_3$ the path length at the previous time step.
With these we arrive at
\begin{multline}
\left[\begin{smallmatrix}
x_1 \\ x_2 \\ x_3 
\end{smallmatrix}\right]_{k+1}
=
\left[\begin{smallmatrix}
1 & 1.50625(1 - \mathrm{e}^{-0.66397T}) & 0  \\ 
0 & \mathrm{e}^{-0.66397T} & 0  \\ 
1 & 0 & 0 
\end{smallmatrix}\right]
\left[\begin{smallmatrix}
x_1 \\ x_2 \\ x_3 
\end{smallmatrix} \right]_k 
\\  +
\left[\begin{smallmatrix}
1.50625T + 2.26879(\mathrm{e}^{-0.66397T}) \\ 
1.50625(1 - \mathrm{e}^{-0.66397T}) \\ 0 
\end{smallmatrix}\right]
u_k 
\label{surge-discrete-time}
\end{multline}
Similarly, the output for this discrete-time system is defined as the difference $\Delta S$ of path lengths between consecutive steps
\begin{equation}
\label{incremental-length}
\Delta S_{k+1} = 1.4162
\left[\begin{smallmatrix}
1 & 0 & -1 
\end{smallmatrix}\right]
\left[\begin{smallmatrix}
x_1 \\ x_2 \\ x_3 
\end{smallmatrix}\right]_{k+1} 
\end{equation}

\begin{figure}[h]
    \centering
    \includegraphics[width=0.7\columnwidth]{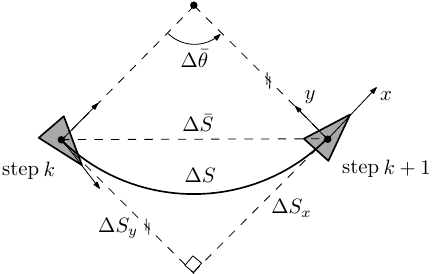}
    \caption{Geometric relation between surge model output and sensor measurements.}
    \label{fig:measurement}
\end{figure}

The challenge here, however, is that $\Delta S$ is not  provided as a sensor measurement.
Nonetheless, this quantity can be derived directly through geometric conditions that depend explicitly on compass and GPS measurements.
Without excessive loss, therefore, we approximate the incremental path length, $\Delta S$, assumed at the scale of the time step to be adequately captured by a circular arc with ZOH over $r_k$, with the distance between the vehicle at consecutive steps, $\Delta \bar{S}$, and express the latter as 
\begin{equation}
    \Delta S_{k+1} \approx \Delta \bar{S}_{k+1} = \sqrt{{\Delta S_x}^2 + {\Delta S_y}^2}
    \label{measurement_eq}
\end{equation}
where $\Delta S_x$ and $\Delta S_y$ can be computed from GPS measurement differences after projection to the body frame at time step $k+1$ (see Fig.~\ref{fig:measurement}).

A remaining challenge, however, is that during deployment and due to the vehicle's control architecture and interface, neither the input thrust $u$ in \eqref{surge model} nor the input moment $r$ are directly known.
This is where retrospective cost input estimation comes in to provide real-time estimates of $u_k$ and $r_k$ through an adaptive estimation process.

\subsection{Adaptive Input Estimation} \label{AIE}
This section provides a mathematical overview of the \ac{AIE} technique~\cite{verma2024real}. 
Consider the linear discrete-time system
\begin{align*}
    x_{k+1} &= A \,x_k + B u_k \\
    y_k &= C\,x_k + V
\end{align*}
where $x_k \in \mathbb {R}^n$ is the system state, $u_k \in \mathbb{R}$ is the control input (assumed unknown), and $V \in \mathbb{R}$ is zero-mean Gaussian sensor noise. 
Matrices $A \in \mathbb{R}^{n \times n}$, $B \in \mathbb{R}^{n \times 1}$, and $C \in \mathbb{R}^{1 \times n}$ are  known. 
The goal of \ac{AIE} is to estimate simultaneously both $u_k$ and $x_k$.
\ac{AIE} is practically comprised of an input estimation subsystem, and a Kalman filter \citep{thacker1998tutorial}. 
\subsubsection{Input estimation subsystem}
To obtain the estimated input $\hat{u}_k$, we construct the input estimation subsystem of order $n_e>1$ as follows
\begin{equation}
    \hat{u}_k = \sum_{i = 1}^{n_e} M_{i,k}\,\hat{u}_{k-i} + \sum_{i = 1}^{n_e} N_{i,k}\,z_{k-i},
    \label{subsystem}
\end{equation} 
where the $M_{i,k} \in \mathbb{R}$, $N_{i,k} \in \mathbb{R}$, 
and $z_k$ is the residual in the prediction step of Kalman filter. 
Subsystem \eqref{subsystem} can be written in form of 
\begin{equation}
    \hat{u}_k = \Phi_k\,\theta_k,
    \label{inpust_est}
\end{equation}
where the regressor matrix $\Phi_k$ is defined as
\begin{equation}
    \Phi_k \triangleq [\hat{u}_{k-1} \cdots \hat{u}_{k-n_e}\; z_k \cdots z_{k-n_e}] \in \mathbb{R}^{1 \times l_\theta}
\end{equation}
the coefficient vector $\theta_k $ is 
\begin{equation}
    \theta_k \triangleq [M_{1,k} \cdots M_{n_e, k}  \; N_{0,k} \cdots N_{n_e,k}] \in \mathbb{R}^{l_\theta}
    \label{coef_vector}
\end{equation}
with $l_\theta \triangleq 2n_e + 1$. 
The order of this subsystem, $n_e$, must be chosen large enough to properly develop the internal model for the input estimation. 
The objective now becomes to update the coefficient vector $\theta_k$, in order to derive an estimate control input. 
To do so, we first define the backward-shift operator for a discrete signal $Y_k$ as 
\begin{align*}
\bm{q}^{-1}\;Y(k) = Y(k-1) 
\end{align*}
and thus we define the following filtered signals
\begin{align*}
    \Phi_{f,k} &\triangleq  G_{f,k}(\bm{q}^{-1})\,\Phi(k-i) 
\\
\hat{u}_{f,k} &\triangleq  G_{f,k}(\bm{q}^{-1})\,\hat{u}(k-i) 
\end{align*}.
By defining $G_{f,k}(\mathbf{q}^{-1}) \triangleq \sum_{i = 1}^{n_f} \mathbf{q}^{-i}H_i(k)$, with $n_f \ge 1$ being the window length of filter, the filtered signals can be written as
\begin{align*}
    \Phi_{f,k} =& \sum_{i = 1}^{n_f}H_i(k)\,\Phi(k-i) 
\\
    \hat{u}_{f,k} =& \sum_{i = 1}^{n_f}H_i(k)\,\hat{u}(k-i)
\end{align*}
where now $H_i(k)$ is defined as
\begin{equation*}
H_i(k) \triangleq  
    \begin{cases}
          CB & k  \ge i = 1 \\
          C\Bar{A}_{k-1}\cdots \Bar{A}_{k-(i-1)}B   
          & k \ge i \ge 2\\
          0 & i > k
    \end{cases}
\end{equation*}
with $\Bar{A} \triangleq A(I + K_{k}C)$, and $K_{k}$ being the Kalman filter gain that is included in the update step of the state filtering process.
To find the coefficient vector $\theta_k$, we construct an optimization problem; more specifically a retrospective optimization problem where the coefficient vector will denote the optimization variable. 
To this end, define the \textit{retrospective variable} as
\begin{equation*}
    z_{r,k}(\hat{\theta}) \triangleq z_k - (\hat{u}_{f,k} - \Phi_{f,k}\hat{\theta})
\end{equation*}
where now the $\hat{\theta}$ represents the optimization argument. 
Next, define the retrospective cost function
\begin{equation*}
    J_k(\hat{\theta}) \!\triangleq\! 
    (\hat{\theta} - \theta_0)^\intercal \!R_\theta (\hat{\theta} - \theta_0)
    +
    \sum_{i=0}^k \!R_z z^2_{r,i}(\hat{\theta}) + R_d(\Phi_i\hat{\theta})^2
\end{equation*}
where $R_z \in (0, \infty)$, $R_d \in (0, \infty)$ are scalar optimization gains, and $R_{\theta} \in \mathbb{R}^{l_{\theta} \times l_{\theta}}$ is a positive definite gain matrix. 
Note that the regularization term $(\hat{\theta} - \theta_0)^\intercal R_\theta(\hat{\theta} - \theta_0)$ weighs the initial estimate and ensures that the $\theta_{k+1}$ has a unique global minimizer \citep{islam2019recursive}. 
Define now $P_0 \triangleq R_{\theta}^{-1}$. 
Then for all $k\ge 1$, the unique global minimizer $\theta_{k+1}$, is given by the  \ac{RLS} update 
\begin{align}
    P_{k+1}  = & P_k - P_k \tilde{\Phi}_k^\intercal \Gamma_k \tilde{\Phi}_k P_k 
\notag
\\
\theta_{k+1} = & \theta_k - P_k \tilde{\Phi}_k^\intercal \Gamma_k (\tilde{z}_k + \tilde{\Phi}_k \theta_k)
    \label{estimated_theta}
\end{align}
where 
\begin{align*}
    \Gamma_k \triangleq &  \big( \tilde{R}^{-1} + \tilde{\Phi}_k P_k \tilde{\Phi}_k^\intercal \big)^{-1}
    &
    \tilde{\Phi}_k \triangleq &
    \begin{bmatrix}
        \Phi_{f,k} \\
        \Phi_k
    \end{bmatrix}
\\
    \tilde{z}_k \triangleq &
    \begin{bmatrix}
        z_k - \hat{d}_{f,k} \\
        0
    \end{bmatrix} 
    &
    \tilde{R} \triangleq &
    \begin{bmatrix}
        R_z & 0 \\
        0 & R_d
    \end{bmatrix}
\end{align*}
By using \eqref{estimated_theta} and replacing the $k+1$ with $k$ in \eqref{inpust_est} we derive the estimated input. We choose $\theta_0 = 0$ which implies $\hat{u}_0 = 0$.

\begin{remark}
    To properly implement the \ac{AIE} algorithm,  it is essential to first specify all the hyperparameters $n_e$, $n_f$, $R_z$, $R_d$, $R_{\theta}$, typically done empirically through trial and error. 
\end{remark}

\begin{remark}
This subsystem is highly sensitive to hyperparameter variations, and small changes in them can either yield the desired results or result in significantly high values for the estimated input.
\end{remark}
\begin{remark}
    Large values of the hyperparameter $n_f$ or $n_e$ do not necessary lead to better filtering of the signals or better estimate control input. 
    On the contrary, they might cause divergence of the input estimate.
\end{remark}

\section{Validation}\label{validation}

This section presents simulation results and numerical analysis that supports the theoretical predictions on (a) estimation of the deterministic control input as it is presented in Section \ref{AIE}, and (b) the effect of the estimated input to the Kalman filter to estimate the state of the Jaiabot. 
The approach is applied to the problem of estimating the states of surge \eqref{surge-discrete-time}--\eqref{incremental-length} and heading model \eqref{heading-discrete-time}--\eqref{incremental-heading}. 
To achieve this, the control inputs for both the surge and heading dynamics must first be estimated, because the operator has no direct knowledge of the thrust and torque inputs.
(The relationship between operator commands and thrust/torque inputs can be empirically established via hydrodynamic experiments.)
For comparison reasons, a standard Kalman filter is also implemented with the assumption that the nominal operator control input is the one that is implemented by the Jaiabot.

Part of the implementation of the algorithm for estimation of control inputs and the states of the two systems, is the process of tuning all the hyperparameters. 
In our tests, we set the following values.
For the \textit{surge dynamics}: $n_e = 4$, $n_f = 8$, $R_z = 1$, $R_d = 50$, $R_\theta = 10^{-0.01}I_9$. 
For the \textit{heading dynamics}: $n_e = 3$, $n_f = 4$, $R_z = 1$, $R_d = 0.1$, $R_\theta = 10^{-2}I_7$.

The algorithms used experimental data collected in Lake Allure, PA, where the Jaiabot was deployed in a series of turning maneuvers (see Fig.~\ref{fig:trajectory}).

\subsection{Results for Surge Dynamics}

\begin{figure}[h!]
    \centering
    \includegraphics[width=\columnwidth]{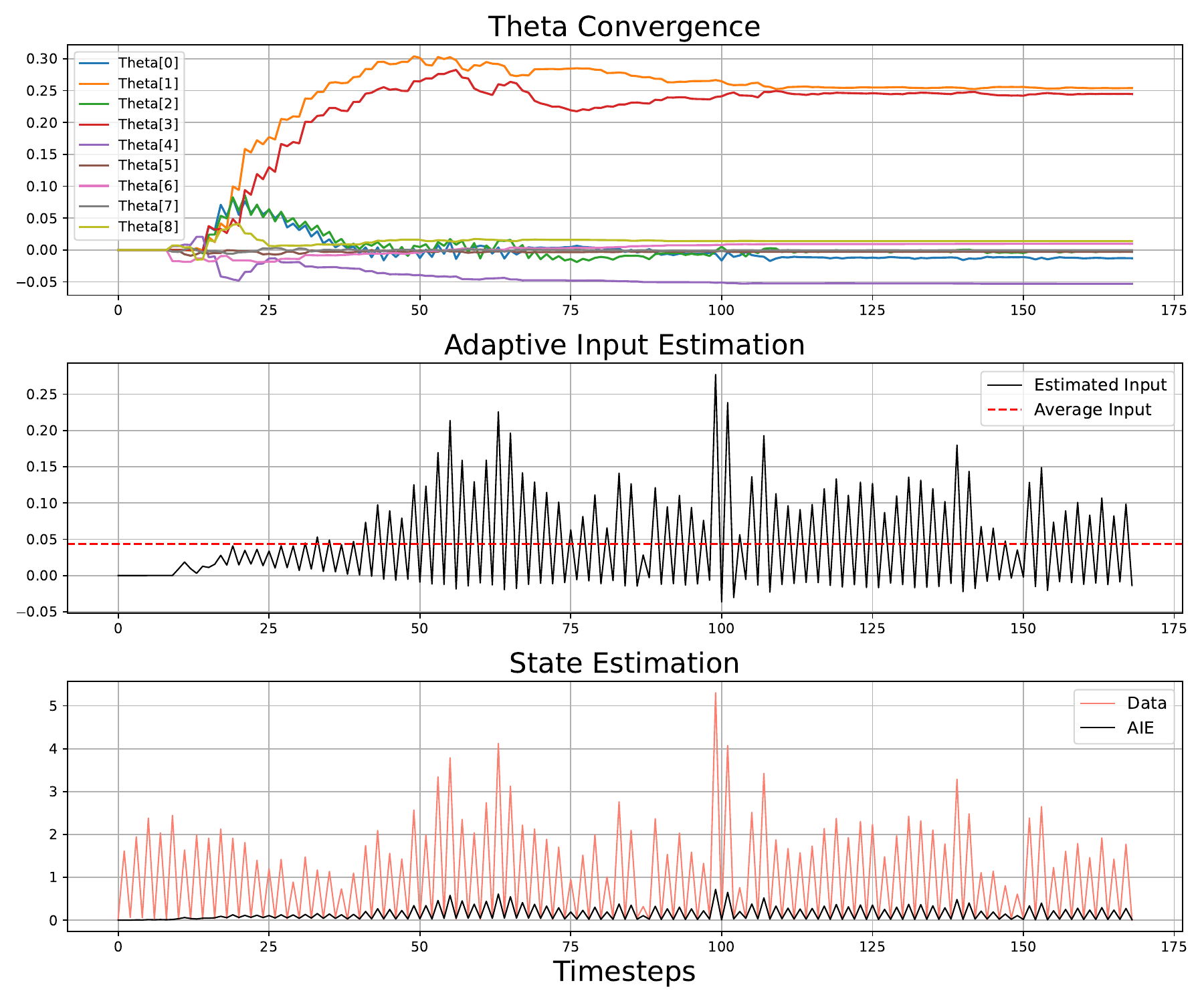}
    \caption{Adaptive input estimation results for surge.}
    \label{fig:surge_AIE}
\end{figure}

First, the \ac{AIE} algorithm has been implemented for the surge dynamics.
Figure~\ref{fig:surge_AIE} gives a comprehensive view of the results obtained from \ac{AIE}. 
As expected, the coefficient vector $\theta_k$ in \eqref{coef_vector} converges after approximately 100 iterations (Fig.~\ref{fig:surge_AIE}, top). 
Nine coefficients $(l_{\theta} = 2n_e +1)$ to achieve proper convergence for this subsystem. 
Figure~\ref{fig:surge_AIE}, middle, illustrates the estimated input for the surge dynamics. 
The dashed red line shows the average of the values that the input takes. 
It can be observed that the control input fluctuates within a small range $([-0.05, 0.30])$, an observation that also aligns with the behavior of the estimated state, shown in Fig.~\ref{fig:surge_AIE}, bottom. 
Note here that the output (and measurement) for the surge dynamics represents the length of the chord of the arc along which the Jaiabot is moving. 
Hence, small values in this context indicate small steps along the motion path.

\begin{figure}[h!]
    \centering
    \includegraphics[width=\columnwidth]{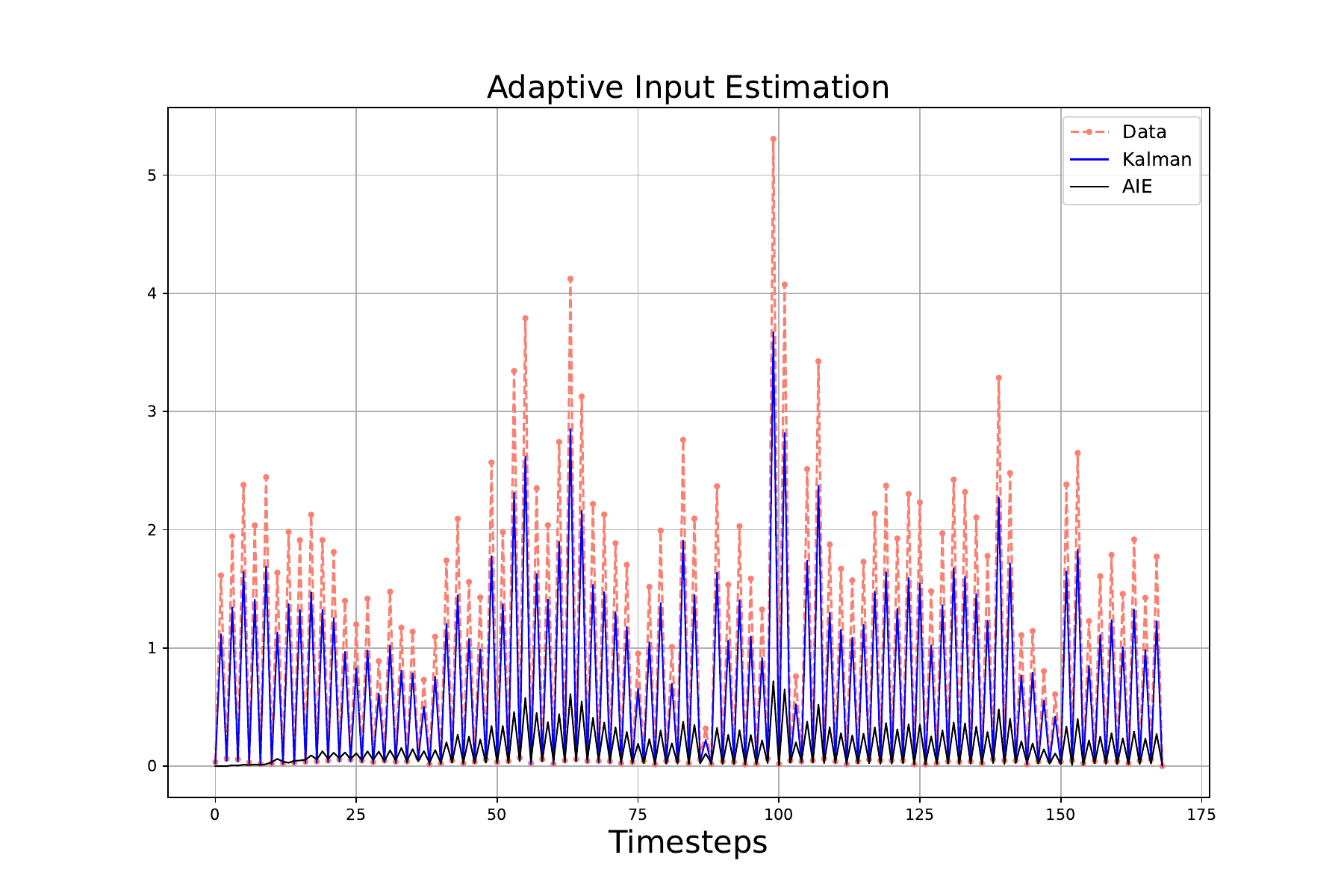}
    \caption{Comparison of the estimated state for surge against experimental data, for the Kalman filter and \ac{AIE}.}
    \label{fig:comparison_surge}
\end{figure}

On the other hand, Fig. \ref{fig:comparison_surge} displays a comparison of state estimation results. 
The red dashed curve represents measurements, the blue curve shows the state estimate obtained from the Kalman filter that uses a nominal (step) control input set at $u_k = 1$, and the black curve illustrates the output of  \ac{AIE}. 
It is seen that \ac{AIE} provides significantly more effective noise filtering compared to the standalone Kalman filter. 

\subsection{Results for heading dynamics}

\begin{figure}[h!]
    \centering
    \includegraphics[width=\columnwidth]{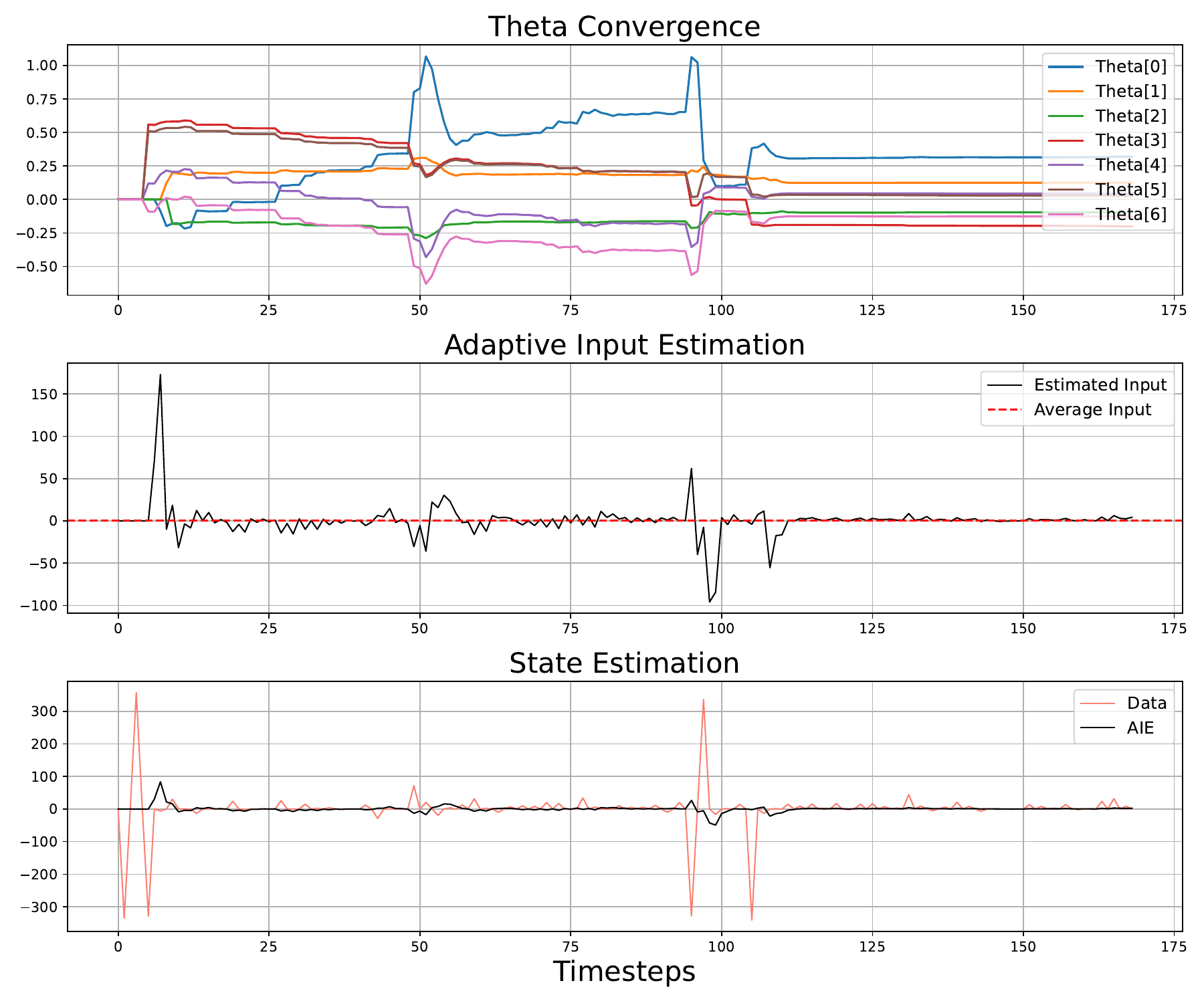}
    \caption{Adaptive input estimation results for heading.}
    \label{fig:heading_AIE}
\end{figure}

In this subsection we report on the \ac{AIE} implementation for the heading dynamics.
Figure~\ref{fig:heading_AIE} summarizes the findings. 
Note that for this model the measurement is the heading (angle) which is obtained directly from the \ac{AUV} sensors. 
Similarly to the \ac{AIE} implementation for the surge dynamics, the coefficient vector $\theta_k$ in \eqref{coef_vector} converges after approximately 100 iterations (Fig.~\ref{fig:heading_AIE}, top). 
For heading estimation, only 7 coefficients are needed to achieve proper convergence. 
Figure~\ref{fig:heading_AIE}, middle, shows the estimated input for the heading dynamics. 
The dashed red curve represents the average of the values that the input takes over the whole period of application. 
In this execution, it can be observed that the control input fluctuates within a range $([-100, 170])$, in correlation with the behavior of the estimated state,  shown in Fig.~\ref{fig:heading_AIE}, bottom, which depicts the estimated state. 
As seen, during instances of significant fluctuations, such as within the range $[80, 110]$, the black line representing the estimation effectively filters out the initial noisy behavior. 

\begin{figure}[h!]
    \centering
    \includegraphics[width=\columnwidth]{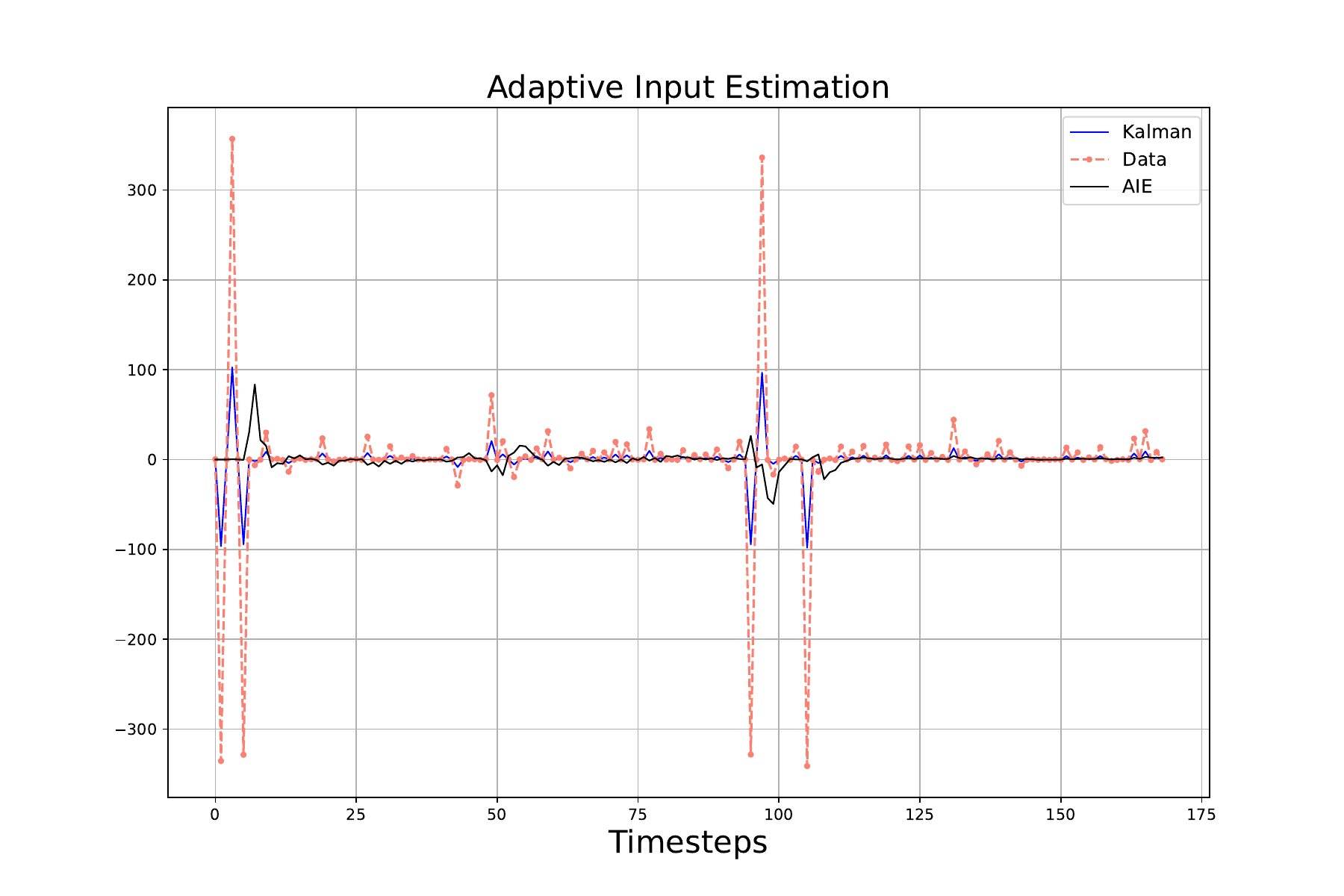}
    \caption{Comparison of the estimated state for heading against experimental data, for the Kalman filter and \ac{AIE}.}
    \label{fig:compasiron_heading}
\end{figure}

Lastly, Fig. \ref{fig:compasiron_heading} displays a comparative analysis of the state estimation results. 
In Fig. \ref{fig:compasiron_heading}, the red dashed line represent system measurements, while the blue lines show the estimated states obtained from the Kalman filter, with the nominal control input set to $u_k = 1$. 
The black line shows the estimated state from the \ac{AIE} algorithm. 
It can be observed that while the Kalman filter provides a good estimate for the heading, it struggles to properly filter the process during significant fluctuations without accurate knowledge of the system input. 
One the other hand, \ac{AIE} framework demonstrates improved performance in scenarios involving large fluctuations, providing more robust and accurate estimates compared to the standalone Kalman filter.

Application of \ac{AIE} suggests that the DC gain of the surge dynamics \eqref{surge-discrete-time} is $4.68$ while that for the heading dynamics \eqref{heading-discrete-time} is $0.125$.

\subsection{Trajectory reconstruction based on estimates}

\begin{figure}[h!]
    \centering
    \includegraphics[width=\columnwidth]{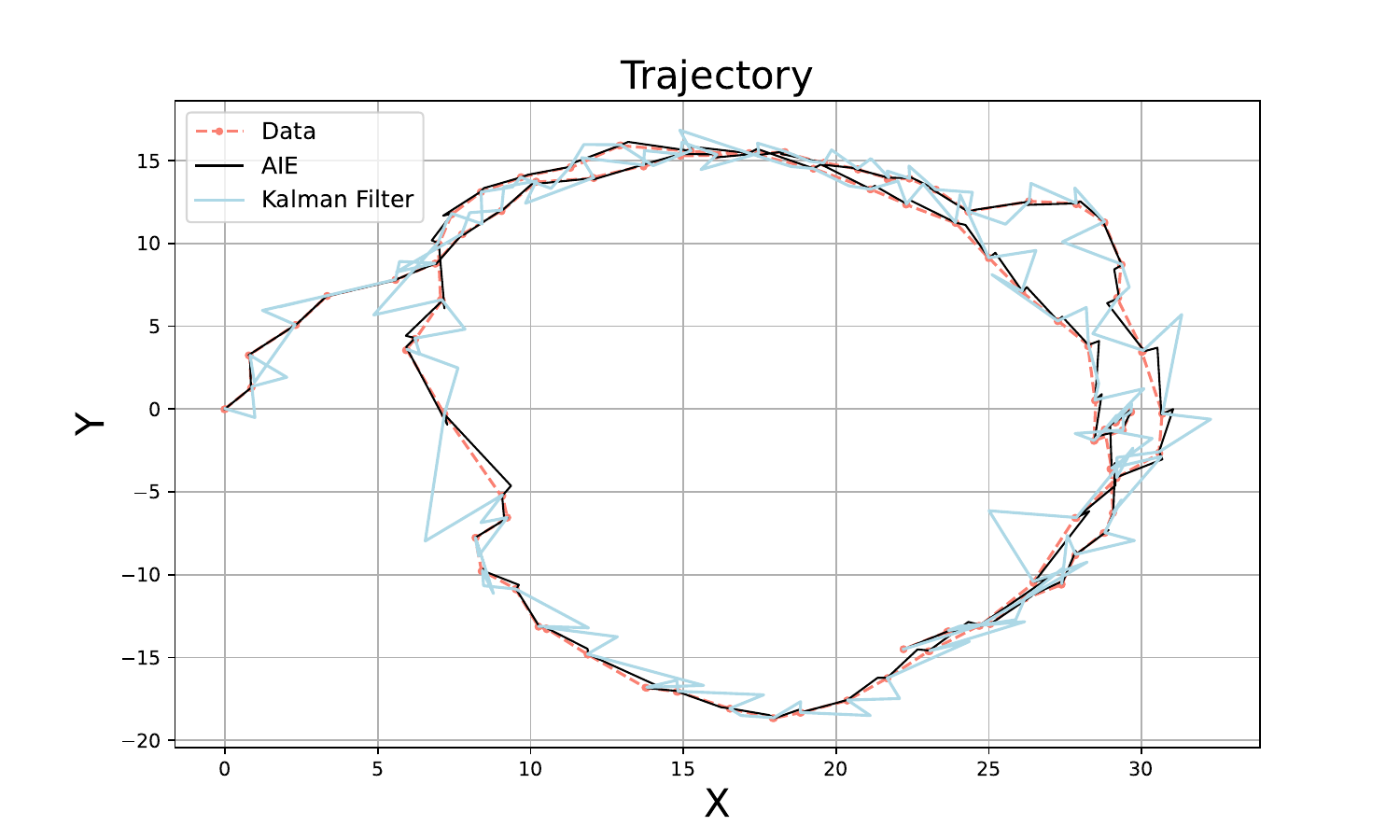}
    \caption{Comparison of the trajectory reconstruction based on the Kalman filter and the \ac{AIE} relative to experimental data.}
    \label{fig:trajectory}
\end{figure}

This section makes a combined spatial comparative analysis between \ac{AIE} and standalone Kalman filter, as the \ac{ASV} motion path is reconstructed based on the corresponding estimates and compared to experimental (GPS) data.
In both cases (\ac{AIE} and standalone Kalman filter) the trajectory is reconstructed as
\begin{align*}
    X_{k+1} = &X_k + \Delta S \, \sin\left(\frac{\Delta \theta}{2}\right)  
    \\
    Y_{k+1} = & Y_k + \Delta S \, \cos \left( \frac{\Delta \theta}{2} \right) 
\end{align*}
where $X_k$ and $Y_k$ are the north-south and east-west coordinates of the \ac{ASV} at time step $k$.
Here, $\Delta \theta$ is derived from the heading dynamics, either using the \ac{AIE} or the Kalman process, while the $\Delta S$ is estimated similarly. 
Figure~\ref{fig:trajectory} depicts the results of this reconstruction for the \ac{ASV} trajectory. 
It can be observed that \ac{AIE} outperforms the Kalman filter since it is adjusting the control input to be suitable for estimation of the state. 
On the other hand, Kalman filter presents big fluctuations along the trajectory, primarily due to the control input that was selected from the user, which does not accurately reflect the actual internal control input implemented by the \ac{ASV}.

\section{Conclusion}\label{conc}

The proper choice of the control input in a system is of the essence for an effective control of a dynamical system. 
In cases where the control input is unknown, it becomes necessary to estimate it \emph{in conjunction} to the state of the system. 
This paper implements this concept for the first time on an \ac{ASV} to generate filtered pose (position \& orientation) estimates using linear models for its surge and heading dynamics, identified independently based on experimental data. 
Results indicate promise for \ac{AIE} as a method for providing accurate and consistent real-time state estimates for an \ac{ASV}, the actual control inputs of which are opaque to its operator. 

\backmatter





\bmhead{Acknowledgements}
The authors thank Chanaka Bandara for collecting and sharing the \ac{ASV} data at Lake Allure PA.
Funding for this work is provided in part through the USGS Next Generation Water Observing System (NGWOS) Research and Development program via award  \#G23AC00149-00 and by NSF through award \#2234869.

\bibliography{FarosJournal.bib}


\end{document}